\documentclass[vecphys]{svmult}

\usepackage{makeidx}         
\usepackage{graphicx}        
\usepackage{multicol}        
\usepackage[bottom]{footmisc}

\makeindex             

\usepackage{epsfig} 

\usepackage{amssymb}
\usepackage{subfigure}
\usepackage{wrapfig}

\def\J{$J/\psi$}

\def\C{c{\bar c}}

\def\q{q{\bar q}}
\def\Q{Q{\bar Q}}
\def\e{\epsilon}

\def\be{\begin{equation}}
\def\ee{\end{equation}}

\def\be{\begin{equation}}
\def\ee{\end{equation}}

\def\lsim{\raise0.3ex\hbox{$<$\kern-0.75em\raise-1.1ex\hbox{$\sim$}}}
\def\gsim{\raise0.3ex\hbox{$>$\kern-0.75em\raise-1.1ex\hbox{$\sim$}}}

\def\NP{{ Nucl.\ Phys.\ }}
\def\PL{{ Phys.\ Lett.\ }}
\def\PR{{ Phys.\ Rev.\ }}

\def\PRL{{ Phys.\ Rev.\ Lett.\ }}

\def\ZP{{ Z.\ Phys.\ }}
\def\EP{{ Europ.\ Phys.\ J.\ }}

\title{\textbf{An Introduction to the Spectral Analysis of the QGP}}
\author{P.\ P.\ Bhaduri$^1$, P.\ Hegde$^2$, H.\ Satz$^3$ and P.\ Tribedy$^1$}
\institute{$^1$VECC, 1/AF Bidhan Nagar, Kolkata-700 064, India. \\
          $^2$Department of Physics and Astronomy, SUNY, Stony Brook, 
          NY 11794-3800, USA. \\
          $^3$Fakult\"at f\"ur Physik, Universit\"at Bielefeld, 
          D-33501 Bielefeld, Germany.}
\date{2nd February, 2008}
\begin{document}
\maketitle
\abstract{This is an introduction to the study of the in-medium
behavior of quarkonia and its application to the quark-gluon plasma
search in high energy nuclear collisions.}

\section{What are quarkonia?}

The bound states of a heavy quark and its anti-quark which are stable with 
respect to strong decay into open charm or bottom are collectively called  
\emph{quarkonia}. We denote by $Q$ either of the heavy quarks, charm ($c$) 
or bottom ($b$); the corresponding bound states are known as \emph{charmonia} 
or \emph{bottomonia}, respectively.

\medskip

Among the vector (spin-one) charmonium states, the lightest (ground state) 
is the famous $J/\psi$; the excited states are the $\chi_c$ and the $\psi'$. 
For the bottom quark, the lightest quarkonium is the $\Upsilon$, while the 
excited states include the $\chi_b$, $\Upsilon'$, $\chi_b'$ and the 
$\Upsilon''$. The stability of the $c\bar c/b \bar b$ quarkonium states 
implies that their masses satisfy $M_{c\bar{c}} < 2 M_D$ and $M_{b\bar{b}} 
< 2 M_B$, where $D = c\bar{u}$ and $B = b\bar{u}$ are the corresponding 
``open'' mesons.
A specific characteristic of quarkonia is their small size. While the 
typical hadron radius is $\sim 1$~fm, the radii of charmonia and bottomonia 
range from $0.1-0.3$ fm, as we shall see.

\medskip

Since $c$ and $b$ quarks are very heavy ($m_Q \ll \Lambda_{QCD} \sim 200$~MeV),
the binding of the $Q\bar{Q}$ system may be treated non-relativistically. 
The governing equation is the non-relativistic Schr\"odinger equation, 
\begin{equation}
-\frac{1}{m}\left\{\nabla^2 (r) +V(r) \right\} \Psi_i(r) 
= (M_i - 2m)\Psi_i(r),
\label{eq:Scheq}
\end{equation}
where $\Psi(r)$ denotes the wavefunction of the system, $r$ the
quark-antiquark separation, and $m$ the
quark mass\footnote{We work in the center-of-mass system, with a reduced mass
$m/2$, so that we have $-\nabla^2/m$ instead of the usual $-\nabla^2/2m$.}.
Since Eq.~(\ref{eq:Scheq}) is a nonrelativistic description of the binding, 
the total rest mass must be subtracted from the masses $M_i$ of the bound 
states. Once we find the eigenvalues $M_i$ of the system, we can also define 
the ``binding energy'' $\Delta E$ of each quarkonium state, 
$\Delta E = 2 M_{D,B} - M_i$.

\medskip

Lattice and spectroscopic studies suggest for the potential $V(x)$
the form \cite{Cornell}
\begin{equation}
V(r) = \sigma r -\frac{\alpha}{r},
\label{eq:Cornell}
\end{equation}
generally known as the ``Cornell potential''. It is spherically symmetric, 
and consists of two parts. The linearly rising part represents the 
confining force, given in terms of the \emph{string tension} $\sigma$; 
lattice studies put its value at around $0.2$ $({\rm GeV})^2$.
The second part is an effective Coulomb potential,
including transverse string oscillations; string theory suggests
$\alpha = \pi/12$.

\medskip

Having solved the Schr\"odinger equation, we may determine the bound-state 
radii through
\be
\langle r_i^2 \rangle = {
\int d^3 r r^2 |\Psi_i(r)|^2 \over \int d^3 r |\Psi_i(r)|^2}. 
\label{rad}
\ee
A fair estimate can already be obtained by means of a semi-classical
formulation. The energy of the system is then given by
\begin{equation}
E = {p^2 \over m}  + V(r),
\end{equation}
and from the uncertainty relation we have $pr \simeq c$; the constant
$c$ can be fixed by requiring the correct \J~mass, giving $c\simeq 1.25$. 
Minimizing the energy determines the lowest bound state radius $r_0$,
\begin{equation}
\sigma +\frac{\alpha}{r_0^2} = \frac{3}{mr_0^3}.
\label{semi}
\end{equation}
With $\sigma \simeq 0.2 (\rm{GeV})^2$ and $\alpha \simeq \pi/12 $, together
with $m_c\simeq 1.3$ GeV, we obtain a \J~size ($\Q$ separation, i.e., twice
the radius) of about 0.5 
fm. For the $\alpha'=0$ value, we have $r_0 \sim (1/m\sigma)^{1/3} 
\approx 0.3$~fm; on the other hand, for $\sigma=0$, we get 
$r_0 \sim (1/m \alpha) \approx 0.6$~fm. 
We thus see that a major contribution to the radius comes from the string 
tension. At $T=0$, the radius of the $J/\psi$ is thus to a considerable
extent still determined by the confining part of the potential. 
We summarize some of the characteristics of the 
spin-averaged quarkonia in Table~\ref{tab:quarkonia} \cite{HSjpg}. 

\medskip

\begin{table}[t]
\centering
\begin{tabular}{|c|c|c|c|c|c|c|c|c|}
\hline 
state & $J/\psi$ & $\chi_{c}$ & $\psi'$ & $\Upsilon$ & $\chi_{b}$ & $\Upsilon'$ & $\chi_{b}'$ & $\Upsilon''$\tabularnewline
\hline
\hline 
mass (GeV) & 3.10 & 3.53 & 3.68 & 9.46 & 9.99 & 10.02 & 10.36 & 10.36\tabularnewline
\hline 
$\Delta E$ (GeV) & 0.64 & 0.20 & 0.05 & 1.10 & 0.67 & 0.54 & 0.31 & 0.20\tabularnewline
\hline 
radius (fm) & 0.25 & 0.36 & 0.45 & 0.14 & 0.22 & 0.28 & 0.34 & 0.39\tabularnewline
\hline
\end{tabular}
\vspace*{0.3cm}
\caption{Masses, binding energies and radii of lowest $c\bar{c}$ and 
$b\bar{b}$ bound states \cite{HSjpg}.}
\label{tab:quarkonia}
\vspace*{-0.3cm}
\end{table}

\medskip

Next, we turn to the question of the dissociation and decay of 
quarkonia. We have already noted that these mesons cannot decay 
via strong channels because their masses are smaller than the open 
thresholds. It is also known that quarkonia do not dissociate 
significantly in nuclear collisions; we shall discuss this in greater 
detail in section~\ref{sec:quarkonium-dissociation}.

\medskip

How then do quarkonia dissociate? Three mechanisms have been identified, 
corresponding to the behavior for $T=0$, $0 < T < T_c$ and $T \geq T_c$, 
where $T_c$ is the critical temperature of deconfinement. We shall consider 
each of them in turn.

\subsection{String-breaking}

\begin{figure}[h]
\centerline{\psfig{file=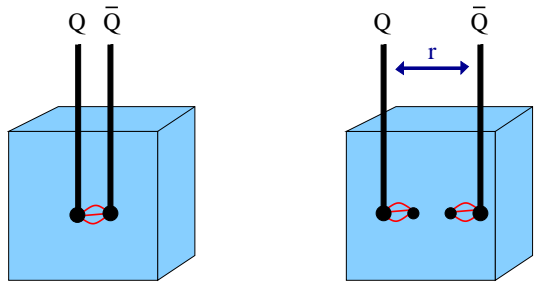,width=7cm}}
\caption{Cartoon of string-breaking}
\label{str-br}
\end{figure}

The potential in Eq.~\ref{eq:Cornell} is correct only in the limit 
$M_Q \to \infty$. If light quarks exist in the theory, then the string 
connecting the heavy quarks can break as soon as the 
overall energy in the system 
is greater than $2 M_D$ or $2 M_B$, depending on whether $Q = c$ or $b$ 
(Fig.~\ref{str-br}). Light quark-antiquark pairs appear at the broken 
ends of the string, and new ``heavy-light'' $Q\bar{q}$ or $q\bar{Q}$ mesons 
are formed. This behavior of the quark potential has been observed in 
lattice studies with dynamical quarks, as we shall show below.

\medskip

\medskip

\begin{figure}[h]
\centerline{\psfig{file=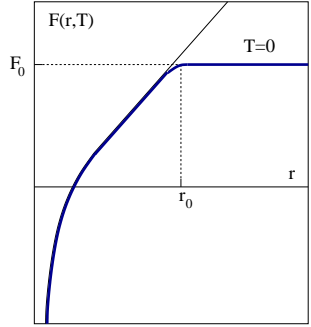,width=4cm}}
\caption{String breaking energy as function of $\Q$ separation}
\label{Fbreak}
\end{figure}

We may estimate the string-breaking energy $F_0$. For the charm quark, 
$F_0 = 2(M_D - m_c) \simeq 1.2$~GeV, while for the bottom quark, 
$F_0 = 2(M_B - m_b) \simeq 1.2$~GeV. From this, we deduce $r_0 = 
(1.2~{\rm GeV})/\sigma \simeq 1.5$~fm. That this value is the same for 
both quark species leads us to conclude that the energy required for
string breaking is a property of the vacuum itself, as a medium at $T=0$,
containing virtual $\q$ pairs which are brought on-shell by the field
between the heavy quarks. The effect of string breaking on the Cornell
potential is shown in Fig.\ \ref{Fbreak}.

\medskip

\subsection{Recombination}

\begin{figure}[htb]
\centering
\includegraphics[width=0.65\textwidth]{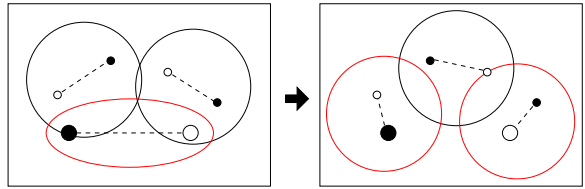}
\caption{A schematic view of recombination}
\label{fig:recomb}
\end{figure}

In nuclear collisions not sufficiently energetic to create a quark-gluon 
plasma, there will nevertheless be abundant hadron production. These newly
formed light hadrons can through a switch in bonding (recombination) 
turn a $\Q$ meson into two heavy-light mesons. This mechanism is schematically
depicted in Fig.~\ref{fig:recomb}: when two or more hadrons overlap, their 
quarks can recouple form new pairs.

\medskip

\begin{figure}[htb]
\centering
\includegraphics[width=0.35\textwidth]{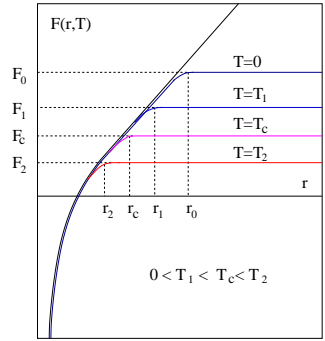}
\caption{Schematic dependence of the string-breaking radius with 
temperature}
\label{fig:str-br-T}
\end{figure}

If the temperature is increased, the hadron density also increases, and 
this in turn increases the recombination probability. As a consequence,
the distance up to which the heavy quarks still bind also becomes shorter,
and the potential will break earlier (see Fig.\ \ref{fig:str-br-T}. We thus 
have something like ``effective screening'', even though all color 
charges are still bound.

\begin{figure}[htb]
\centering
\subfigure{
\epsfig{file=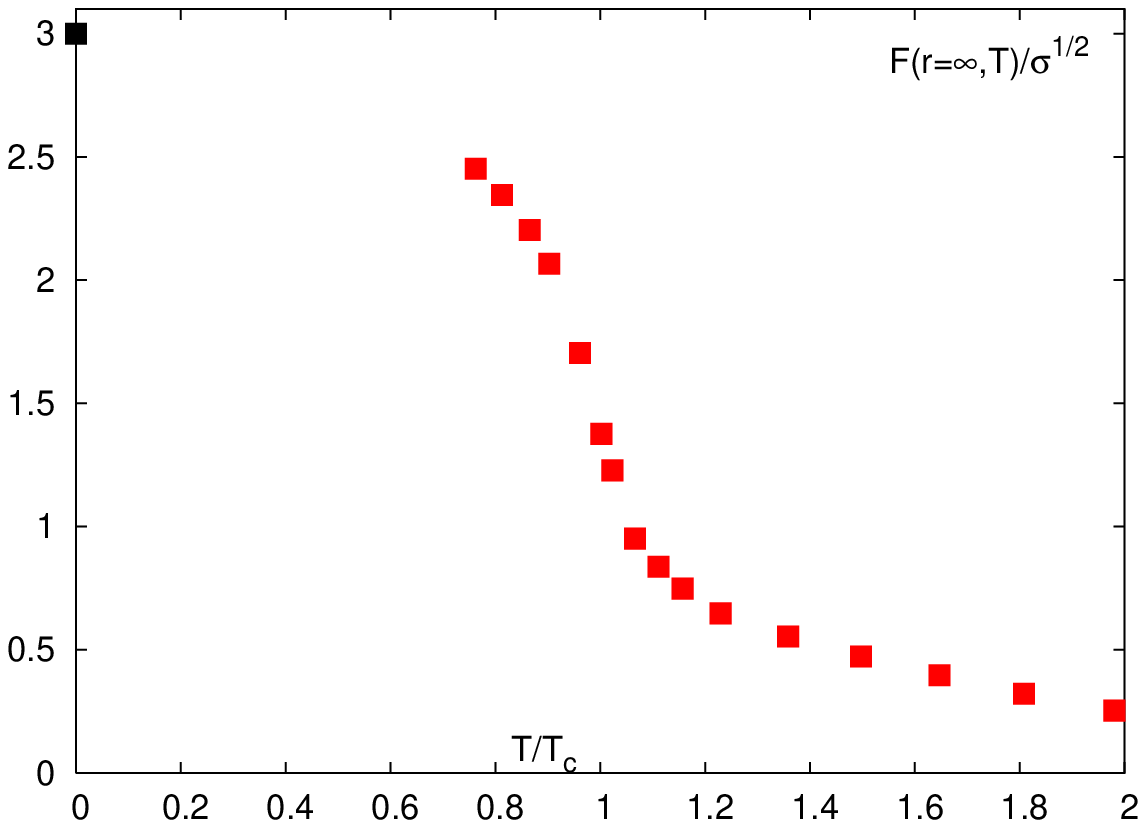,width=5.5cm}
}
\subfigure{
\epsfig{file=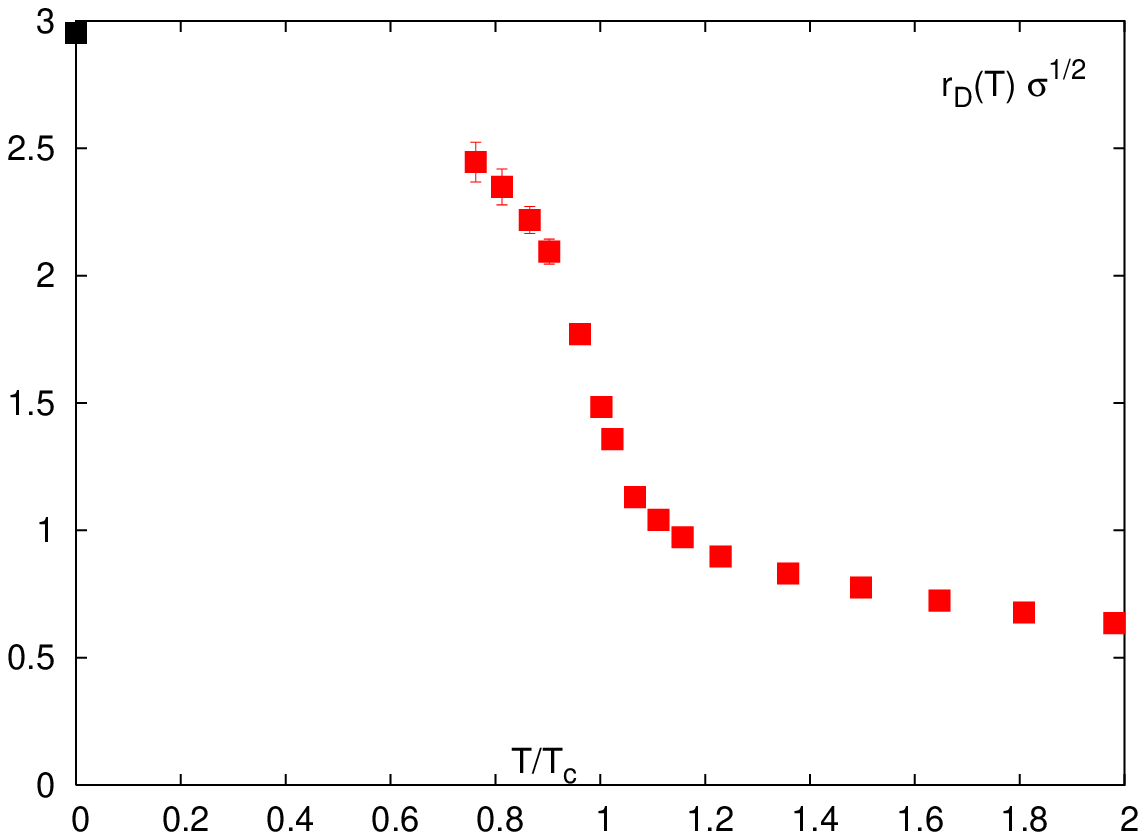,width=5.5cm}
}
\caption{Lattice results for free energy and screening radius
as function of $T$.}
\label{near}
\end{figure}

What happens as we get close to $T_c$? The density of produced hadrons
will then increase strongly, and  lattice studies show that in accord with
our picture, both the free energy and the string-breaking radius $r_T$ 
decrease rapidly near $T_c$, as shown in Fig.\ \ref{near}.

\subsection{Color screening}

Above $T = T_c$, we have a medium of unbound color charges, and 
an entirely different mechanism takes over. At all temperatures $T$ 
above zero, quarks and gluons are \emph{screened}, just 
as electric charges experience Debye screening 
in an electromagnetic plasma.
This screening occurs with a characteristic radius, which we denote 
by $r_D$. It decreases with increasing temperature, as the medium
increases in density.
Deconfinement is expected to occur when this radius becomes comparable to 
the average hadron size of 1~fm. Then a given quark can no longer see its
former partner in a hadron; instead, it sees many other quarks and 
antiquarks and therefore can move around freely, without encountering
any confinement limit, since it is never 1 fm away from an antiquark.

\medskip

\begin{figure}[t]
\centering
\includegraphics[width=0.4\textwidth, height=0.2\textheight]{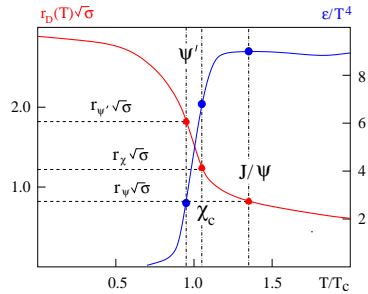}
\caption{Quarkonium dissociation as ``thermometer'' 
for the quark-gluon plasma.}
\label{fig:debye-L}
\end{figure}

We would like to use the behavior of the \J~to probe if a quark-gluon
plasma was formed in the collision \cite{matsui-satz}.
However, as we have seen, the $J/\psi$ and its heavier counterparts have 
smaller radii than the usual hadrons. Thus, charmonia and bottomonia may 
be expected to survive beyond the QGP phase transition up to some 
higher temperature, at which they will become dissociated. Thus,
if we know their sizes as well as the behaviour of $r_D$ as a function of 
$T$, we can use their dissociation points to determine the temperature 
and the energy density $\e$ of the QGP medium \cite{asi}, as
illustrated in Fig.~\ref{fig:debye-L}. 

\section{Studying charmonium dissociation}
We now turn to the question of how to determine quantitatively the
quarkonium dissociation points in a quark-gluon plasma.
Two different approaches were used to address this problem.
\begin{itemize}
\item{Solve the Schr\"odinger equation with a temperature-dependent 
potential $V(r,T)$, or}
\item{calculate the quarkonium spectrum directly in finite temperature
lattice QCD.}
\end{itemize}
We shall look at each of these approaches in turn.

\subsection{Potential models for quarkonium dissociation}

\subsubsection{The Schwinger model \cite{KMS}}

One generalizes the Cornell potential, eq.~(\ref{eq:Cornell}), to non-zero 
temperature in the form
\begin{equation}
V(r,T) = \sigma r\left\{\frac{1-e^{-\mu r}}{\mu r}\right\} 
-\frac{\alpha}{r}e^{-\mu r}.
\label{eq:Sch-potl}
\end{equation}
The $T$-dependence of the above equation is in the ``screening mass'' 
$\mu(T) = 1/r_D(T)$. Eq.~(\ref{eq:Sch-potl}) gives the correct
zero-temperature limit, eq.~(\ref{eq:Cornell}), for $\mu(T)\to 0$ 
as $T \to 0$. 

\medskip

To determine the dissociation points, one solves the Schr\"odinger 
equation and determines the bound-state energies $M_i(\mu)$. With 
increasing temperature, the bound state $i$ disappears at some
$\mu = \mu_i$. One then uses the temperature dependence of the
screening mass from lattice estimates, $\mu(T) \simeq 4~T$, to 
determine the $T_i$. The result of this model is 

\begin{itemize}
\item{the $\psi'$ and $\chi_c$ become dissociated around $T\simeq T_c$,}
\item{the $J/\psi$ survives up to about $T \simeq 1.2 T_c$.}
\end{itemize}

In both cases, at the dissociation point the binding energy vanishes,
while the binding radius diverges. 

\subsubsection{Lattice potential models \cite{D-P-S1,SZ,
Wong,Alberico,D-K-K-S}}

Alternatively, one may use lattice results for the temperature dependence 
of the potential felt by a static quark-antiquark pair to determine
the needed potential. The static $\Q$ studies
start from the partition function $Z$, which is related to the 
free energy by $Z = \exp (-\beta F)$; this in turn gives the
thermodynamic potentials 
$$
F = U - TS 
$$
$$
S = -\left(\frac{\partial F}{\partial T}\right)_V 
$$
\be
U(r,T)= F(r,T) - T\left(\frac{\partial F(r,T)}{\partial T}\right)_V
\ee
Assuming that the internal energy $U(r,T)$ provides the 
temperature dependence of the heavy quark potential, we
we use results from $N_f = 2$ lattice QCD and solve the Schr\"odinger 
equation. The results obtained from such studies indicate that

\begin{itemize}
\item{the $\psi'$ and $\chi_c$ are dissociated around a temperature 
$T \simeq 1.1 T_c$,}
\item{$J/\psi$ survives up to a temperature $T \simeq 2 ~T_c$.}
\end{itemize}

Comparing these results to the ones from the Schwinger model, we see 
that while there is agreement in the case of the higher excited states,
lattice potential models predict a considerably higher dissociation 
temperature for the $J/\psi$. The reason for this is that 
the internal energy $U(r,T)$ 
leads to much stronger binding than the Schwinger model potential. 

\medskip

It should be noted here that there still is some ambiguity as to whether 
$U$ or $F$ is the correct potential to be used in the Schr\"odinger
equation. Hence there exist approaches with potentials of the form 
$aU+(1-a)F$, with $0 \leq a \leq 1$.
Such potentials tend to reduce binding and lower the dissociation 
temperature as $a$ is decreased. 

\subsection{Lattice studies of charmonium survival} 

\begin{figure}[!t]
\centering
\includegraphics[width=0.6\textwidth]{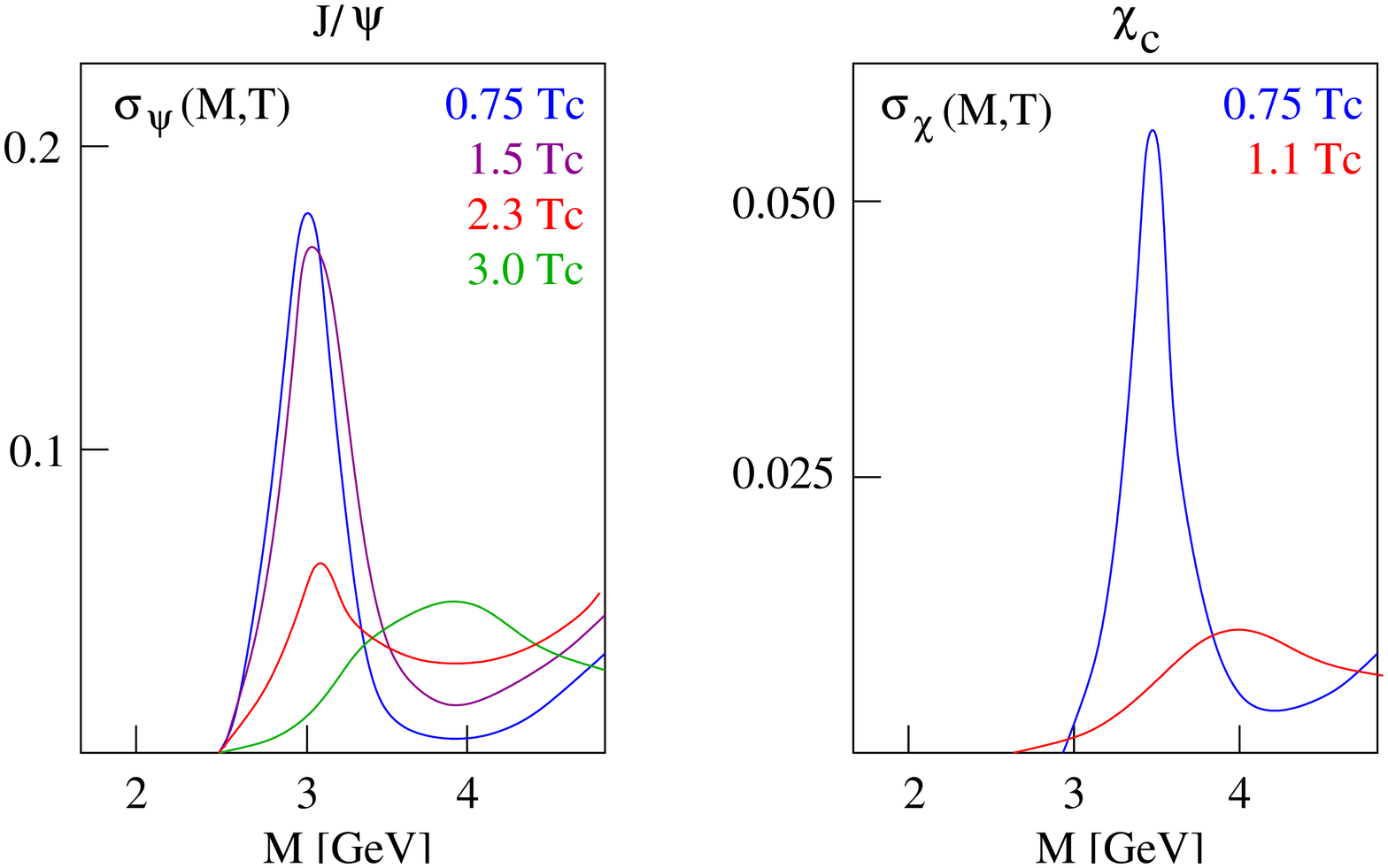}
\caption{Schematic view of lattice results for charmonium
dissociation.}
\label{lattice-sigma}
\end{figure}

The ideal way to resolve the above ambiguity would be to calculate the 
$c\bar{c}$ spectrum directly on the lattice, and this is indeed what 
lattice studies aim to do \cite{Umeda,Asakawa,Datta,Iida,Jacovac,
Skullerud}. More specifically, they calculate the 
$c\bar{c}$ spectrum $\sigma(\omega,T)$ in the appropriate quantum channel, 
as a function of the temperature $T$ and the $c\bar{c}$ energy $\omega$. 
Bound-states show up as resonances in a plot of $\sigma$ versus $\omega$. 
By performing simulations at different temperatures, one can determine 
the temperature at which a particular peak disappears~i.e.~a bound-state 
dissolves. A schematic illustration is shown in Fig.\ \ref{lattice-sigma}.
The results presently indicate that 
\begin{itemize}
\item{$\chi_c$ is dissociated for $T\geq 1.1T_c$.}
\item{$J/\psi$ persists upto $1.5 < T/T_c < 2.3$.}
\end{itemize}
Thus, on the basis of lattice studies, the following picture emerges: 
The higher excited states dissociate around $T = T_c$, while the 
$J/\psi$ survives up to much higher temperature, in accord with
the potential model studies based on the internal energy $U(r,T)$.

\medskip

There is, however, a caveat to these calculations. The discretization 
introduced by the lattice limits the resolution of the peak. Lattice 
methods are thus useful in determining the position and to some extent
the amplitude of the peaks, but determining the peak \emph{widths} 
remains a challenge, nor is it easy to study the spectrum in the 
continuum region ($\omega > 4$~GeV).

%
\section {Dynamics of quarkonium dissociation}
\label{sec:quarkonium-dissociation}
We have seen in the previous discussion that the $J/ \psi$, the vector 
ground state of charmonium family, is very tightly bound. Its binding 
energy i.e. the energy difference between $J/\psi$ mass and open charm 
threshold, $\Delta E_{J/\psi}$, is considerably larger than the typical 
non-perturbative hadronic scale $\Lambda_{QCD}$, 
\begin{equation}
\Delta E_{J/\psi} = 2M_D-M_{J/\psi} = 0.6~\rm{GeV} \gg \Lambda_{QCD}. 
\sim 0.2~\rm{GeV}
\label{eq.3.1}
\end{equation}

Consequently the size of $J/\psi$ is much smaller than that of typical hadron,
\begin{equation}
r_{J/\psi} ~\sim~ 0.25~\rm{fm} \ll \Lambda^{-1}_{QCD}~\sim~1~\rm{fm}   
\label{eq.3.2}
\end{equation}
We now want to consider by what kind of dynamical interaction such 
a state can be dissociated.
Because of the small spatial size, the $J/\psi$ can only be resolved by a 
sufficiently hard probe. Moreover, because of its high binding energy, 
only a sufficiently energetic projectile can break the binding. The 
previous study of global medium effects had led to the conclusion that 
only a hot deconfined medium, consisting of colored quarks and gluons,
is capable of dissociating the charmonium vector ground state. We now
want to study this on a microscopic level.

\medskip

In a deconfined medium, the constituents are unbound partons, whereas 
in a confined medium the constituents are hadrons. Such thermal 
hadrons are incapable of causing collisional dissociation 
of $J/\psi$. Let us illustrate this point.

\medskip

\begin{figure}[htb]
\centering
\includegraphics[height=3.5cm]{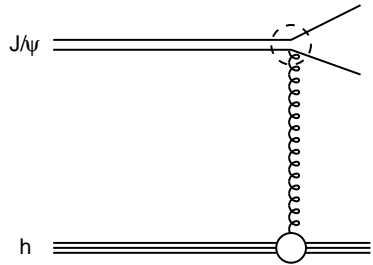}
\caption{Schematic view of the interaction of a normal hadron with a \J.}
\label{3.1}
\end{figure}

Consider the collision of a \J~with a normal hadrorn. Because of the 
small characteristic \J~size, only a hard partonic constituent of the
hadron can see the \J~and interact with it. In other words,
$J/\psi$ collisions with ordinary hadrons  
probe the local partonic structure of these `light' hadrons, not their 
global aspects such as mass, size, or overall momentum. The parton
nature of the interaction is illustrated in Fig.\ \ref{3.1}.

To see the effect of this more quantitatively \cite{KS94}, we take
an ideal pion gas as the confined medium. The momentum 
distribution of the pions at a temperature $T$ follows 
$f(p) \sim \exp(-|p|/ T)$, giving the pions an average momentum 
$\langle p \rangle \sim 3~\!T$. Now the gluon momentum distribution 
inside a hadron as determined by deep inelastic lepton-hadron scattering 
is given by the parton distribution function $g(x)$; here 
$ x = 2k_g/\sqrt s $, with $k_g$ for the gluon momentum, so that
$x$ may be thought of as the fraction of the incident hadron's 
momentum carried by the gluon. For the pionic gluon it takes the form 
\begin{equation}
g(x)~\sim~(1-x)^3.
\label{eq.3.4}
\end{equation}

The resulting average gluon momentum in the hadron thus becomes 
\begin{equation}
\langle k_g \rangle = p_h \cdot \frac{\int_0^1 dx~x~g(x)}
{\int_0^1 dx~g(x)}.
\label{eq:k_av}
\end{equation}
With eq.\ (\ref{eq.3.4}) and $p_h = 3T$ for the momentum of the 
incident hadron, we obtain
\begin{equation}
\langle k_g\rangle_h = \frac{p_h}{5} = {3T\over5}  \leq 0.1~\rm{GeV},
\label{eq.3.5}
\end{equation}
where we have assumed $T < 175$ MeV for the temperature of the hadronic 
medium.
Thus gluons bound inside the hadronic constituents of confined matter 
are much too soft to cause the dissociation of a $J/\psi$. 

\medskip

On the other hand, in a deconfined medium, such as an ideal QGP, the gluons 
are free and distributed according to a thermal distribution 
$f(k_g) \sim \exp(-k_g/T$), which gives
\begin{equation}
\langle k_g \rangle ~\sim~3T
\label{eq.3.7}
\end{equation}
so that for T$~\geq~1.2~T_c \simeq 0.63$ GeV, the 
gluons are hard enough to overcome the $J/\psi$ binding.

\medskip

We have thus noticed that
deconfinement results in a hardening of the relevant gluon momentum 
distributions. 
More generally speaking, the onset of deconfinement will lead to 
parton distribution functions which are different from those for 
free hadrons, as determined by DIS experiments. Since hard gluons 
are needed to resolve and dissociate $J/\psi$'s , one can use $J/\psi$s 
to probe the in-medium gluon hardness and hence the confinement status 
of the medium.

\medskip

This qualitative picture can be made quantitative by short distance 
QCD calculations \cite{KS94,Bhanot}. One has to calculate first the 
cross section for 
gluon dissociation of $J/\psi$, a QCD analogue of the photo-effect. 
This can be carried out using the operator product expansion, which 
is essentially a multipole expansion for the charmonium quark-antiquark 
system. Fig. \ref{3.1} shows the relevant diagram for the calculation 
of inelastic $J/\psi$-hadron cross section. The upper part of the figure 
corresponds to $J/\psi$ dissociation by gluon interaction. The cross 
section for this process has the form 
\begin{equation}
\sigma_{g-J/\psi}~\sim~{1\over m^2_c}{(k/\Delta E_\psi -1)^{3/2}\over 
(k/\Delta E_\psi)^5}
\label {eq.3.9}
\end{equation}
with $\Delta E_{J/\psi} = 2M_D-M_{J/\psi}$.
The corresponding cross section for the hadron dissociation is obtained 
by convoluting this gluon dissociation cross section with the gluon 
distribution function $g(x)$ of the incident hadron. For $J/\psi$-meson 
interactions, this leads to the form 
\begin{equation}
\sigma_{h-J/\psi}~\simeq~\sigma_{\rm{geom}}(1-{\lambda_0\over\lambda})^{5.5}
\label {eq.3.10}
\end{equation}
with $\lambda \simeq (s-M^2_\psi)/M_\psi$ and 
$\lambda_0 \simeq (M_h+\Delta E_\psi)$, where $\sqrt s$ is 
the CMS energy of the $J/\psi$-hadron system.
Here $\sigma_{\rm{geom}} \simeq \pi r^2_{J/\psi} \simeq~ 2$~mb is the 
geometric $J/\psi$ cross section and $M_h$ denotes the mass of the 
incident meson.
\begin{figure}[!tbh]
\centering
\includegraphics[height=6cm,width =5cm,angle=-90]{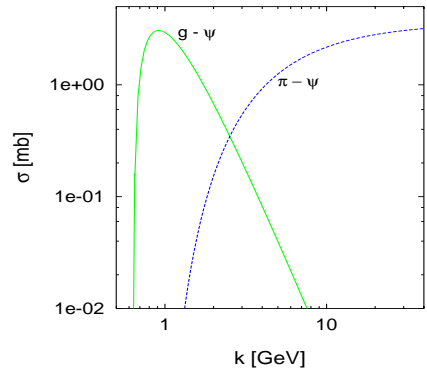}
\caption{Cross sections for \J~dissociation by gluons vs.\ pions}
\label{Fig:3.2}
\end{figure}
Fig.~\ref{Fig:3.2} compares the two dissociation cross sections,
$J/\psi$ dissociation by gluons (gluo-effect) and by pions, 
as a function of projectile momentum $k$ incident on stationary 
$J/\psi$, as given by eqs.~(\ref{eq.3.9}) and (\ref{eq.3.10}). 
The gluon cross section shows the typical photo-effect form,
vanishing until the gluon momentum $k_g$ reaches the binding 
energy $\Delta E_{J/\psi}$; it peaks just a little later 
($ \lambda_g \sim r_{J/\psi}$) and then vanishes again when 
sufficiently hard gluons just pass through the (comparatively larger) 
charmonium bound states ($\lambda_g \ll r_{J/\psi}$). 
In contrast, the $J/\psi$-hadron inelastic cross section 
remains negligibly small until rather high hadron momenta (3-4 GeV). 
In a thermal medium such momenta corresponds to temperatures of more 
than 1 GeV. In other words, in a confined medium in the temperature 
range of the order of a few hundred MeV the $J/\psi$ should survive, but 
it should become dissociated in a hot deconfined medium.
Confined media in the temperature range of few hundred MeV 
are thus essentially transparent to a $J/\psi$, while a deconfined 
medium of the same temperature is opaque to $J/\psi$'s and very 
efficiently dissociates them.

\section{Quarkonium production in nuclear collisions}   

The aim of ultra-relativistic nuclear collisions is to study color 
deconfinement and the resulting quark-gluon plasma in the laboratory. 
We want to use quarkonia produced in the collision as a 
probe to study the medium produced in the collision. Both 
the quarkonium states and the medium to be probed require 
a {\it`finite formation time'}, so we have to look at the evolution 
aspects in both cases.
Let us first consider the issue of charmonium production in 
hadron-hadron collisions and then turn to nuclear targets.
\subsection{Quarkonium production in hadronic collisions}

\begin{figure}[htb]
\centering
\includegraphics[height=4.5cm]{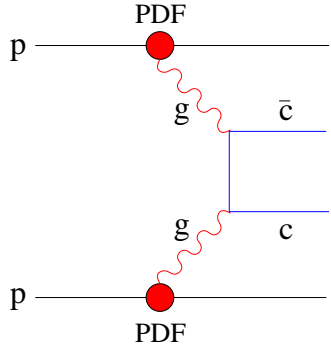}
\caption{Lowest-order Feynman diagram for $c\bar{c}$ production 
through gluon fusion.}
\label{Fig:4.1}
\end{figure}

Quarkonium production in hadron-hadron collisions occurs in three stages. 
The first stage is the production of $c\overline c$ pair.
Because of the large quark mass ($m_c\sim1.3$ GeV ) this process can be 
treated as a hard process and is well described by perturbative QCD. 
A parton from the projectile interacts with one from the target; 
the (non-perturbative) parton distributions within the hadrons are 
determined empirically in other reactions, e.g., by deep inelastic 
lepton-hadron scattering. At high energy the process of  
c$\overline c$ production dominantly occurs by gluon fusion, 
$gg\rightarrow c \bar c$ (see Fig.\ \ref{Fig:4.1}).
\medskip

\medskip

The $c \overline c$ in general is in a color octet state. It has to 
neutralize its color in order to leave the interaction zone and form 
a physical resonance like $J/\psi$ or $\psi'$. In the second stage, 
color neutralization occurs by interaction with the surrounding color 
field. This results finally in the third stage of a physical bound state. 
Both the second and third stages are non-perturbative in nature.

\medskip

On a fundamental theoretical level, color neutralization is not yet fully 
understood, but there are several models, color singlet \cite{Baier-R}, 
color octet \cite{CO}, and color evaporation \cite{CE}. The 
color evaporation model provides a particularly
simple and experimentally well-supported phenomenological 
approach. In the evaporation process, the $c\bar c$ can either combine 
with light quarks to form open charm mesons ($D$ and $\bar D$) or 
bind with each other to form a hidden charm (charmonium) state. A fixed 
fraction of the subthreshold $c\bar c$ production is used in charmonium 
production. The basic quantity in this picture is the total sub-threshold 
charm cross section $S_{c \bar c}$, obtained by integrating the 
perturbative $c\bar c$ production cross section $\sigma$ over the mass 
window from $2m_c$ to $2m_D$. Since at high energy, the dominant part 
of $S_{c\bar c}$ comes from gluon fusion (Fig.~\ref {Fig:4.1}), we can write 
\begin{equation}
S_{c \overline c}(s)~\simeq~\int^{2m_D}_{2m_c}d\hat{s} \int dx_1~dx_2~g_{p}(x_1)~g_{t}(x_2)~ \sigma(\hat{s})~ \delta(\hat{s}-x_1x_2s),
\label{eq.3.11}
\end{equation}
with $g_{p}(x)$ and $g_{t}(x)$ denoting the gluon densities and
$x_1$ and $x_2$ the fractional momenta of the gluons from projectile 
and target, respectively; $\sigma$ is the $gg\rightarrow c\bar c$ 
cross section.

\medskip

As mentioned, the basic assumption of the color evaporation model 
is that the production cross section for any particular charmonium state 
is a fixed fraction of the subthreshold charm cross section,
\begin{equation}
\sigma_i(s)= f_i~S_{c \bar c}(s)
\label{eq.3.12}
\end{equation}     
where $f_i$ is an energy-independent constant to be determined empirically. 
It follows that the energy dependence of the production cross section for 
any charmonium state is predicted to be that of the perturbatively calculated 
sub-threshold charm cross section. As a further consequence the production 
ratios of different charmonium states 
\begin{equation}
{\sigma_i(s)\over \sigma_j(s)} = {f_i\over f_j} = \rm {constant}
\label{eq.3.13}
\end{equation}
must be energy independent.
Both these predictions have been  compared in detail to charmonium 
hadro-production data over a wide range of energies \cite{quarko}. 
They are found 
to be well supported, both in the energy dependence of the cross sections 
and in the constancy of the relative species abundances.

\medskip

Before turning to the topic of quarkonium production in hadron-nucleus  
collisions, let us consider the relevant time scales for the $J/\psi$ 
formation.

\medskip

The formation of a $c\bar c$ pair requires a time $\tau_{c\overline c} =
1 / 2m_c = 0.05$ fm. The produced $c \bar c$ pair is in a 
color-octet state. To form a physical resonance state, it has to 
neutralize its color. The color-octet model \cite{CO}
proposes that the color-octet 
$c \bar c$ combines with a soft collinear gluon to from a color-singlet 
$(c\bar c-g)$ state. After a short relaxation time $\tau_8$ this 
pre-resonance $(c\bar c-g)$ turns into physical resonance by absorbing 
the accompanying gluon, with similar formation processes for the other 
resonances, such as  
$\chi_c$ and $\psi^{\prime}$ production. The color-octet model 
encounters difficulties if the collinear gluons are treated 
perturbatively, indicating once more that color neutralization 
seems to require non-perturbative elements. 
However it does provide a conceptual basis for the evolution 
of the formation process.
\begin{figure}[htb]
\centerline{\epsfig{file=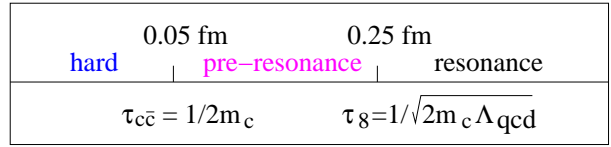,width=8cm}}
\caption{Evolution of \J~formation in a hadron-hadron collision}
\label{scales}
\end{figure}

\medskip

The color neutralization time $\tau_8$ of the pre-resonant state can 
be estimated by the lowest momentum possible for the confined gluons 
$\tau_8 \simeq \left(2m_c\Lambda_{QCD}\right)^{-1/2} \simeq 0.25$ fm.
The resulting scales of $J/\psi$ formation are illustrated in 
Fig(\ref{scales}). The formation time for the actual physical 
ground state $J/\psi$ is presumably somewhat larger than $\tau_8$, 
although $r_{J/\psi}~\simeq~\tau_8$ , the heavy $c$ quarks move 
non-relativistically. For the larger higher excited states , 
the formation time will then be correspondingly still larger. 
\subsection{Quarkonium production in pA and AA collisions}

Let us now turn to nuclear collisions. Both in p-A and A-A interactions
there will be pre-resonant absorption in nuclear matter. In 
nucleus-nucleus collisions, however, there can be in addition
a substantial amount of a produced {\it``secondary medium''}, and 
testing this medium is in fact our main objective. 

\medskip

The creation of the medium and production of the probe lead to two 
distinct formation scales. In p-A collision 
there is no formation time for the medium, so that such collisions 
provide a tool to probe charmonium production, evolution and absorption 
in confined matter.

\medskip

Nuclear effects can arise in all the evolution stages of $J/\psi$ 
production, and a number of different phenomena have to be taken into 
account.
\begin{itemize}
\item{The presence of other nucleons in the nucleus can modify the 
initial state parton distribution functions, which enter in the 
perturbative $c\bar c$ production process, as shown in Fig. \ref{Fig:4.1}. 
This can lead to a decrease (shadowing) or to an increase (antishadowing) 
of the production rate.}
\item{Once it is produced, the $c \bar c$ pair in its evolution
will traverse the nuclear matter; it can suffer absorption 
both in the pre-resonance as well as in the resonance stage, caused 
by successive interactions with the target nucleons.}
\end{itemize}

\medskip

\begin{figure}[htb]
\centering
\includegraphics[height=3.5cm]{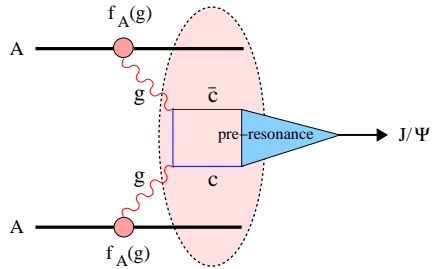}
\caption{\J~production in a nuclear medium.}
\label{Fig:4.4}
\end{figure}

\medskip

Hence $J/\psi$ production in a nuclear medium is modified as compared 
to hadronic collisions. The modification occurs before any QGP formation 
and is thus independent of the effects due to a deconfined medium 
having free quarks and gluons. If we want to use $J/\psi $ 
production and its suppression in a nuclear collision as a potential 
signature of the QGP formation, all normal nuclear effects must first
be taken into account. Only then can charmonium suppression serve as 
a probe to test the confining status  of the produced 
``secondary medium'' in nuclear collisions.

\medskip

So an essential question is how to account for the effects of the nuclear 
medium initially present on the production. The basis for this,
both in theory and in experiment, is the measurement of
dilepton, open charm and charmonium production in p-A or d-A collisions. 
These collisions thus provide a crucial tool to understand 
quarkonium production in nuclear collisions. 

\medskip

The procedure to be used for such studies is the following:
\begin{itemize}
\item{The initial state parton distribution functions in nuclear matter 
are determined by open charm and dilepton production in p-A/d-A collisions 
in the relevant kinematic region.}
\item{The Glauber model then is used to determine the pre-resonance absorption 
of the $J/\psi$ and $\psi^\prime$ by the target nucleon in p-A/d-A in 
the relevant kinematic region.}
\end{itemize}
It is thus clear that p-A or d-A collision experiments are an absolutely
essential tool for the analyis of 
quarkonium production in nuclear collisions.

\subsection{Sequential quarkonium suppression}

\begin{figure}[!tbh]
\centering
\includegraphics[height=4cm]{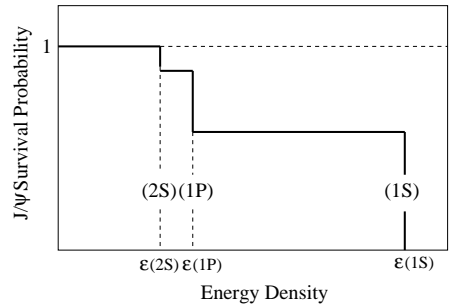}
\caption{Sequential $J/\psi$ suppression} 
\label{Fig:4.5}
\end{figure}

There is a further important and, as it turns out, crucial feature 
observed in \J~hadroproduction. 
The $J/\psi$ actually measured in hadron-hadron collisions are not all 
directly produced $1S$ charmonium states; rather, they have three distinct 
origins. 
About 60$\%$ of them are indeed directly produced 1S charmonium states,
but the rest are feed-down from higher excited states. 
About 30$\%$ come from the decay $\chi_c(1P)\rightarrow J/\psi~+$ anything 
and the remaining 10$\%$ from $\psi^{\prime}(2S)\rightarrow J/\psi~+$ 
anything. In both cases, the decay widths of the involved higher excited 
states are extremely small (less than 1 MeV), so that their lifetimes 
are very long and the decay occurs long after the interaction.
The presence of any medium in nuclear collisions would 
therefore affect these excited states themselves and not their products,
and we had seen above that excited states are dissociated before the
ground state is.  
This has a direct consequence on the nature of $J/\psi$ suppression by 
deconfinement. In a thermal QCD medium, we should expect that 
with increasing temperature or 
energy density, first the $J/\psi$ originating from $\psi^\prime$ decay 
and then those from $\chi_c$ decay will disappear. Only a considerably 
higher temperature would be able to remove the directly produced $J/\psi$s. 
Such a stepwise onset of suppression with specified threshold temperatures 
is perhaps the most characteristic feature predicted for charmonium 
production in nuclear collisions. It is illustrated schematically in 
Fig.\ \ref{Fig:4.5}, where we have defined the $J/\psi$ survival 
probability to be unity if the production rate suffers only the 
estimated normal nuclear suppression. The generic suppression 
pattern shown here will of course be softened by nuclear profile effects, 
impact parameter uncertainties, etc. On the other hand, this could be 
partially compensated if there is a discontinuous onset of deconfinement 
as a function of energy density of the medium.

We had seen above how to calculate the quarkonium dissociation points which 
specify the temperature and thus also the energy density of the medium, 
thereby serving as a QGP thermometer. 
Potential model studies based on the heavy quark internal energy,
as well as direct lattice QCD calculations gave as dissociation 
temperatures $T \simeq  1.1~ T_c~\rm{for}~\psi^\prime ~{\rm and}~\chi_c$
and $ T \geq  1.5 - 2~ T_c~\rm{for}~J/\psi$.
If this is correct, then the direct $J/\psi$(1S) survives up
to about  $\e\ge~10 - 20~\rm{GeV fm^3}$. Consequently, 
all anomalous suppression observed at SPS and RHIC must be due to the 
dissociation of higher excited states $\chi_c~\rm{and}~\psi^{\prime}$
 \cite{KKS}. 
The suppression onset for this is predicted to lie around $\epsilon~
\simeq~1~\rm{GeV/fm^3}$, and once these are gone, only the unaffected 
$J/\psi$ (1S) production remains. Hence the $J/\psi$ survival probability 
(once normal nuclear effects are taken into account) should be same for 
central Au-Au at RHIC as for central Pb-Pb collisions at SPS.
\subsection{Charmonium regeneration}
In this section we want to investigate the possibility that the medium 
produced in high energy nuclear collision differs from the deconfined 
state of matter studied in finite temperature QCD. The basic idea here 
is that nuclear collisions initially produce more than the thermally 
expected charm and this excess, if it survives, may lead to a new form 
of combinatorial charmonium production at hadronization 
\cite{PBM,Thews,Rapp}.
\medskip

In the QGP argumentation, a crucial aspect was that the charmonia,
once dissociated, can not be recreated at the hadronization stage, 
because of the extremely low thermal abundance of charm quarks
in an equilibrium QGP.
The thermal production rate for a $c\bar c$ pair relative to a 
pair of light quarks is 
\vspace*{-0.2cm}
\begin{eqnarray}
{n_{c \bar c}\over n_{q\bar q}}& \simeq & \exp-(2m_c-2m_q/T_c~) \\ \nonumber
&\simeq& \exp(-2m_c/ T_c) \simeq 3.5\times 10^{-7},  
\label{eq:3.16}
\end{eqnarray} 
with $m_c=1.3$ GeV for charm quark mass and $T_c$ = 175 MeV for the 
transition temperature. The initial charm production in high energy 
hadronic collisions is, however, a hard non-thermal process, and the 
resulting rates calculated from perturbative QCD are considerably larger
than the thermal rate. 
Moreover, in AA interactions the resulting $c /\bar c$ production rate 
grows with the number of binary collision $N_{\rm{coll}}$, while the light 
quark production rate, being soft process, grow as the number of 
participants nucleons, i.e much slower. At high 
collision energies, the initial charm abundance in AA collisions is thus 
very much higher than the thermal value. Now the question is, what happens 
to this in course of the collision evolution?

\medskip
The basic assumption of the regeneration approach is that the initial 
charm excess is maintained throughout subsequent evolution, i.e, the 
initial chemical non-equilibrium will persist up to the hadronization point. 
In charmonium hadroproduction, $J/\psi$ are formed because some of 
the $c \bar c$ pairs produced in a given collision form the corresponding 
bound state. In a collective medium formed through the superposition 
of many nucleon-nucleon (NN) collisions, such as a quark-gluon plasma, 
a $c$ quark from one NN collision can in principle also bind with a $\bar c$ 
from another NN collision (``new'' pairs) to create a $J/\psi$. 
This pairing provides a ``exogamous'' charmonium production mechanism, 
in which the c and $\bar c$ in a charmonium state have different parents, 
in contrast to ``endogamous'' production in pp collision. At sufficiently 
high energies this can lead to an enhancement in $J/ \psi$ production 
in AA collisions compared to a scaled pp rates \cite{PBM,Thews,Rapp},
provided the overall charm 
density is sufficiently high at hadronization and provided the binding 
probability between charm quarks from different sources is large enough.
\medskip

\begin{figure}[!tbh]
\centering
\includegraphics[height =4.5cm]{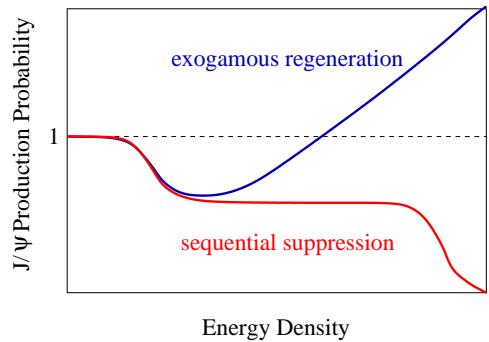} 
\label{Fig:4.6}
\caption{Statistical \J~regeneration vs. sequential \J~suppression}
\end{figure}

Whether or not such enhancement becomes significant depends on two factors. 
On one hand, the initial charm oversaturation must be preserved so that the 
total charm abundance is non-thermal. On the other hand it is necessary 
that the recombination between charm quarks from different parents to 
charmonium ($J/\psi$) is strong enough. Here it is generally assumed that 
the final hadronization occurs according to the available phase space. 
Thus the number of statistically recombined $J/\psi$ has the form 
$N_{J/\psi}~\sim~ N_{c \bar c}^2$, growing quadratically in 
the number of $\C$ pairs. This implies that the hidden 
to open charm ratio, e.g., $ N_{J/\psi} / N_D ~\sim~ N_{c \bar c} 
/ N_h $ increases with energy, in contrast to the energy independent 
form obtained for the fully equilibrated QGP, or to the decrease predicted 
by color screening. The prediction for \J~production by regeneration
is compared in Fig.\ \ref{Fig:4.6} to that from sequential suppression.

\section{Conclusion}

Statistical QCD predicts the existence of a new state of nuclear matter, 
the quark-gluon plasma (QGP), at very high temperatures and/or densities. 
This medium, in contrast to hadronic matter, is capable of dissociating 
quarkonia, so that {\it quarkonium suppression} may be taken 
as a sign of QGP formation in nuclear collisions \cite{matsui-satz}.
Furthermore, different quarkonia dissociate 
at different temperatures; the dissociation pattern thus serves as 
a ``thermometer'' for the QGP. It is therefore important to obtain precise
predictions for these dissociation points, and for this, one can turn to 
either of two approaches, potential models or lattice studies. The former 
have the problem that the results are dependent on the type of potential 
chosen, while the latter so far suffer from the fact that lattice 
spacing and statistics limits the resolution of peak widths in the 
spectrum. It is also not easy to identify the continuum region of the 
spectrum on the lattice. 

\medskip

But what happens in case of relativistic nuclear collisions in the 
laboratory? If there is no regeneration of the dissociated charmonia, 
$J/ \psi$ remain as an external probe, and the sequential suppression 
pattern of the $J/\psi$ can then serve as a tool to determine the 
energy density and the temperature of the produced medium.
On the other hand, if there is \J~production through statistical 
combination of $c$ and $\overline c$ from different collisions,
leading to an overall \J~enhancement, this 
would clearly indicate the thermalization of the produced medium 
on a pre-hadronic level. However, charmonia could then no longer serve 
as a thermometer to charaterize the primordial medium.
Data from LHC, soon to come, will certainly play a decisive role 
in settling the issue.




\begin{thebibliography}{99.}
%
%
%



\bibitem{Cornell} E.\ Eichten et al., \PR D17 (1978) 3090; \PR D 21 (1980)
203.

\bibitem{HSjpg} For a recent review, see H.\ Satz, J.\ Phys.\ G 32 (2006) R25.

\bibitem{matsui-satz} T.~Matsui and H.~Satz, Phys.~Lett.~B, {\bf 178}, 416 (1986).
\bibitem{asi} For a recent review, see H.~Satz, \NP A 783 (2007) 249c.

\bibitem{KMS} F.\ Karsch, M.-T.\ Mehr and H.\ Satz, \ZP C 37 (1988) 

\bibitem{D-P-S1} S.\ Digal, P.\ Petreczky and H.\ Satz, \PL B 514 (2001) 57.

\bibitem{SZ} E.\ Shuryak and I.\ Zahed, \PR D {\bf 70}, 054507 (2004).

\bibitem{Wong} C.-Y.\ Wong, \PR C 72 (2004) 034906;\\ 
C.-Y.\ Wong, hep-ph/0509088;\\
C.-Y.\ Wong, \PR C 76 (2007) 014902.

\bibitem{Alberico} W.\ Alberico et al., \PR D 72 (2005) 114011.

\bibitem{D-K-K-S} S.\ Digal et al., \EP C 43 (2005) 71.
  
\bibitem{Umeda} T.\ Umeda et al., Int.\ J.\ Mod.\ Phys.\ A16 (2001) 
2215.

\bibitem{Asakawa} M.\ Asakawa and T.\ Hatsuda, \PRL 92 (2004).

\bibitem{Datta} S.\ Datta et al., \PR D 69 (2004) 094507

\bibitem{Iida} H.\ Iida et al., PoS LAT2005 (2006) 184.

\bibitem{Jacovac}A.\ Jacovac et al., \PR D 75 (2007) 014506. 

\bibitem{Skullerud} R.\ Morrin et al., PoS LAT2005 (2006) 176;\\
G.\ Aarts et al., \NP A 785 (2007) 198.

\bibitem{KS94} D.\ Kharzeev and H.\ Satz, \PL <b334 (1994) 155.

\bibitem{Bhanot} M.\ E.\ Peskin, \NP B 156 (1979) 365;\\
G.\ Bhanot and M.\ E.\ Peskin, \NP B 156 (1979) 391.

\bibitem{Baier-R} R.\ Baier and R.\ R\"uckl, \ZP C 19 (1983) 251.

\bibitem{CO} G.\ T.\ Bodwin, E.\ Braaten and G.\ P.\ Lepage, \PR D
51 (1995) 1125;\\
E.\ Braaten and S.\ Fleming, \PRL 74 (1995) 3327.

\bibitem{CE} M.\ B.\ Einhorn and S.\ D.\ Ellis, \PR D12 (1975)
2007;\par
H.\ Fritzsch, \PL 67B (1977) 217;\par
M.\ Gl\"uck, J.\ F.\ Owens and E.\ Reya, \PR D17 (1978) 2324; \par
J.\ Babcock, D.\ Sivers and S.\ Wolfram, \PR D18 (1978) 162.

\bibitem{quarko} H.\ Satz and X.-N.\ Wang (Eds.), {\sl Hard Processes 
in Hadronic Interactions}, Int. J. Mod. Phys. A 10 (1995) 2881;

M.\ Mangano et al. (Eds.), {\sl Hard Probes in Heavy-Ion Collisions at the
LHC}, CERN Yellow Report 2004-09, Geneva 2004. 

\bibitem{KKS} D.\ Kharzeev, F.\ Karsch and H.\ Satz, \PL B 637 (2006) 75.

\bibitem{PBM} P.\ Braun-Munzinger and J.\ Stachel, \NP A690
(2001) 119.

\bibitem{Thews} R.\ L.\ Thews et al., \PR C 63 (2001) 054905.

\bibitem{Rapp} L.\ Grandchamp and R.\ Rapp, \NP A 709 (2002) 415.


\end{thebibliography}
\end{document}